# SOME MUSINGS ON GLOBULAR CLUSTER SYSTEMS

Sidney van den Bergh
Dominion Astrophysical Observatory
National Research Council of Canada
5071 West Saanich Road
Victoria, British Columbia
Canada  V9E 2E7
sidney.vandenbergh@nrc.ca

## ABSTRACT

Globular cluster systems exhibit a bewildering variety of characteristics. No single scenario appears to be able to account for the wide range of specific globular cluster frequencies that are observed in galaxies of various types.  The fraction of all star formation that produces massive bound clusters differs from galaxy to galaxy, and (in some cases) appears to vary with time.  Presently available data strongly suggest that the specific cluster forming frequency is highest during violent bursts of star formation.  Globular cluster systems show a wide variety of color (metallicity) distributions, with many luminous galaxies having double (or even multiple) metallicity peaks that were probably produced during distinct episodes of cluster formation.  The bulk of the globulars in the main body of the Galactic halo appear to have formed during a single short-lived burst that took place  ~ 13 Gyr ago.  This was followed by a second, perhaps more extended, burst associated with the formation of the Galactic bulge.  A few metal-rich clusters may also be associated with the formation of the Galactic Thick Disk. Finally many of the clusters beyond $R_{Gc} \simeq 15$ kpc might have been formed in dwarf spheroidal galaxies that, after a few Gyr, were accreted by the main body of the Galactic halo. Some of these outer clusters are younger, and less luminous, than the majority of globular clusters formed at smaller Galactocentric distances.

1. **INTRODUCTION**

All giant galaxies, and quite a few lesser ones, are embedded in huge globular cluster systems. The fact that the Galactic nuclear bulge is surrounded by a swarm of globulars was first noted by Shapley (1918a,b). Subsequently Hubble (1932) discovered that M 31 was also surrounded by a large system of globular clusters. Information, with various degrees of completeness, is presently available for 50 - 100 globular cluster systems (Gebhardt & Kissler-Patig 1999, Kundu 1999, Neilsen & Tsvetanov 1999, and references therein). Some of the systematics of their properties have been discussed by Harris (1991, 1999, 2000), Kissler-Patig (1997, 2000), and by Forbes et al. (1996, 1997a, 2000).

Observations of nearby galaxies (Larsen & Richtler 1999, 2000) show that the fraction of all star formation that takes place in massive clusters differs significantly from object to object. Quiescent galaxies like IC 1613 (van den Bergh 1979, Wyder, Hodge & Cole 2000) seem to have a low specific cluster frequencies, whereas active ones, such as the nearest starburst galaxy NGC 1569, are observed to have high specific cluster frequencies. Furthermore, massive clusters appear to be ubiquitous (Fritze-von Alvensleben 1999) in merger remnants, of which NGC 4038/39 (= The Antennae) is the best known prototype. Contrary to expectations, the clusters in the Antennae appear to have formed with a log-normal mass spectrum. It would be important to obtain independent confirmation of this surprising result. There is presently some tentative evidence which seems to suggest that massive star clusters preferentially form during the most violent and chaotic phases of galaxy evolution. In this connection, it is noted that Larson (1993) had previously speculated that "The formation of a dense bound cluster may be the culminating event in a region of star formation." Alternatively, one might speculate that shocks are required to trigger the formation of such massive star clusters.

Observations of the Large Magellanic Cloud suggest (van den Bergh 1999a) that two major bursts of star and cluster formation have occurred in this galaxy during its lifetime. The first of these appears to have taken place ~ 13 Gyr ago, just after the collapse of the proto-LMC, whereas the second ramped up between 3 and 5 Gyr ago (Butcher 1977). Available evidence strongly suggests that the first spike in the rate of globular cluster formation (Da Costa 1991, Olsen et al. 1998, Johnson et al. 1999) in the Large Cloud occurred when the LMC was being assembled, or when it collapsed to a disk. Of the clusters that formed during the second LMC starburst only the populous young cluster NGC 1866 (van den Bergh 1999b) appears to have a mass which rivals that of the 13 globulars (Suntzeff 1992) that formed during the first burst.

It is presently not understood why this first great burst of globular cluster formation in the Large Cloud was synchronous with that in the main body of the Galactic halo, and with the event that formed the outer halo globular cluster NGC 2419 (Harris et al. 1997). It is noted in passing that all of these globulars that formed more-or-less simultaneously have metallicity [Fe/H] < -1.2 (Rosenberg et al. 1999).[1] What triggered this enormous burst of globular cluster formation in

---

[1]    The globular cluster Fornax No. 4 (Marconi et al. 1999) has [Fe/H] $\simeq$ -2.0, but appears to be a few Gyr younger than the majority of globulars. This shows that some metal-poor globular clusters have ages that are significantly smaller than ~ 13 Gyr.

---

the Galaxy, its halo, and in its nearest companions remains a complete mystery. Possibly, strong shocks played a role in the formation of the first generation of massive clusters (Kumai, Basu & Fugimoto 1993). It is curious that the guillotine which suddenly terminated this great burst of cluster formation in the main body of the Galactic halo does not appear to have operated very effectively in dwarf spheroidal galaxies (van den Bergh 2000). As a result a significant fraction of the globulars associated with dwarf spheroidals (e.g. Fornax 4 and Terzan 7) are a few Gyr younger than bulk of the globulars surrounding the Galaxy and the Large Magellanic Cloud. It is not clear why the initial burst of massive cluster formation occurred in galaxies as small as the Fornax dwarf spheroidal (Buonanno et al. 1998), but not in the more massive Small Magellanic Cloud (Rich et al. 2000).

The Triangulum galaxy M 33 has a luminosity comparable to that of the LMC. It is therefore puzzling that the globular cluster systems of these two objects exhibit such different dynamical properties. Schommer et al. (1991) have shown that the M 33 globulars have the large velocity dispersion that is diagnostic of halo kinematics. On the other hand Schommer et al. (1992) show that the globular clusters in the LMC obey disk kinematics.[2] Why did the M 33 globulars

---

[2] Alcock et al. (2000) note that the LMC RR Lyrae stars also appear to have a disk-like distribution.

---

form during the halo phase of its evolution, whereas the LMC globular clusters were not produced until the Large Cloud had collapsed to a disk? Perhaps an even deeper mystery is provided by the flattened (E3.5-like) globular cluster system that surrounds the giant E0 galaxy M 87 (Kundu et al. 1999). Observations by Cohen (2000) show that this flattened globular cluster system is rotating with a velocity of ~ 300 km s$^{-1}$. How could the flattened, fast-rotating primordial system from which these globular clusters formed have collapsed into a spherical E galaxy? The M 87 globular cluster system has a bimodal color distribution (Kundu et al. 1999, and references therein), indicating that these objects were formed during two distinct evolutionary phases. Surprisingly both the blue subsystem [E(3 ± 0.5)], and the red subsystem [E(4 ± 0.5)] have similarly flattened spatial distributions. It will be a real challenge to theoreticians to derive scenarios that can account for the fact that the Virgo E0 galaxy NGC 4486 (M 87) has a significantly flattened globular cluster system, whereas the Virgo E1 galaxy NGC 4472 (M 49) is embedded in a globular cluster system that appears to be spherical (Harris & Petrie 1978).

It is still not entirely clear why such a large fraction of luminous galaxies are embedded in cluster systems that have bimodal color distributions. This bimodality indicates that many luminous galaxies experienced at least two major epochs of globular cluster formation (Berman & Suchkov 1991). In the case of our own Milky Way system the first of these two eras appears to have coincided with the formation of the Galactic halo, whereas the second may have been associated with the earliest phase of bulge formation. Furthermore the existence of a final phase of globular cluster formation, that might have been associated with a third (Thick Disk) phase of Galactic evolution, is hinted at by a few metal-rich clusters, such as M 71 and 47 Tucanae (van den Bergh 1993, Harris 2000). In fact, the true history of the Galactic globular cluster system may be even more complex, with many of the clusters at $R_{Gc} > 15$ kpc having been formed in dwarf spheroidal galaxies that were later accreted (and destroyed) by the Galaxy. Such accreted globulars tend to be younger, larger, and less-luminous (van den Bergh 2000) than those that had earlier formed in the main body of the Galaxy and its halo.

## 2. THE SPECIFIC GLOBULAR CLUSTER FREQUENCY

### 2.1 Specific frequency differences

The specific cluster frequency S (Harris & van den Bergh 1981) is defined as the number of globular clusters per $M_V = -15$ of parent galaxy light. The single most striking characteristic exhibited by the S values of galaxies is that S(Ellipticals) > S(Spirals). Many years ago van den Bergh (1982a) suggested that this observation militates against the hypothesis that elliptical galaxies were formed from merging spirals. However, more recently Schweizer (1987) has argued that this difference is, in fact, due to the fact that the majority of globular clusters in elliptical galaxies actually formed during the violent collisions between gas-rich ancestral spirals that resulted in the assembly of such luminous elliptical galaxies.

Can mergers increase the <u>relative</u> number of globular clusters enough to account for the observed difference between the S values of spirals and ellipticals? Harris (private communication) emphasizes the fact that collisions will trigger the formation of both new field stars and new clusters. So S will only increase if an above-average fraction of new stars form in clusters. Harris (1991) finds that S ~ 0.5 in Sc/Ir galaxies, S ~ 1 in spirals of types Sa/Sb, S ~ 2.5 in field ellipticals and S ~ 5 for ellipticals in rich clusters. Part of these differences can be explained by evolutionary fading of late-type galaxies, which will increase S over time. However, it is not yet clear if cluster formation during mergers between gas-rich galaxies can account for the entire difference between <S(spirals)> and <S(ellipticals)>. In this connection it is noted (Forbes 1999) that none of the mergers that have been studied so far appears to have increased the number of clusters by more than a factor of two. A remaining problem with the merger scenario is that luminous E galaxies (which are believed to have been assembled via mergers of spirals), and S0 galaxies (that are thought to have mainly evolved from a single parental spiral), appear to have similar mean S values (Harris 1991). For recent discussions of the effects of mergers on globular cluster populations the reader is referred to the monograph Globular Cluster Systems by Ashman & Zepf (1998) and to the excellent critical review by Elmegreen (1999).

A variant of the cluster formation during galaxy merger scenario has been proposed by Smith (1999) who suggests that globular clusters formed when the protogalactic "bits and pieces" of Searle & Zinn (1978) collided. A possible problem with this scenario is that one might expect it to result in a correlation between galaxy mass (and hence the collisional velocities of the fragments) and the masses and luminosities of the clusters produced during these collisions. However, no such correlation is actually observed.

Yet another scenario has, among others, been advanced by Taniguchi, Trentham & Ikeuchi (1999). These authors envision a situation in which globulars are formed in the shocked shell that is created when galactic winds, driven by the massive burst of star formation in a galactic bulge, sweep up metal-poor ambient of infalling gas. Such a picture might account for the metal-poor globular cluster population that has little (or no) net angular momentum. Ikeuchi (private communication) suggests that more than one such shell may be required to account for the different globular cluster populations that are observed in galaxies.

Forbes et al. (2000) have shown that if the Local Group were to merge into a single "elliptical" its globular clusters would have a bimodal color distribution with peaks at [Fe/H] ~ -1.55 and [Fe/H] ~ -0.65. The merged system would have S ~ 1, which is smaller than that of typical ellipticals. The very red integrated color $(B-V)_o \approx 0.96$ (de Vaucouleurs 1961) of M 31 (which contributes most of the luminosity of the Local Group) shows that this value of S will not increase very much if the evolutionary dimming of starlight is taken into account. The E3 galaxy NGC 1700, which appears to have formed from merging spirals ~ 3 Gyr ago (Brown et al. 2000), also has a low specific cluster frequency of S = 1.4.

### 2.2 High S values in cD galaxies

The mean specific globular cluster frequency in early-type galaxies appears to be weakly dependent on parent luminosity (Harris 1991), with both the faint Fornax dwarf spheroidal and very luminous cD/E galaxies having above-average S values.  Furthermore the specific frequency of ellipticals in the Virgo cluster may, on average, be as much as twice that in lesser clusters and in the field. McLaughlin Harris & Hanes (1994) have suggested that the central cD galaxies in the least dynamically evolved clusters of types BM II and BM III tend to exhibit very high S values, whereas central galaxies in the more dynamically evolved clusters of type BM I often have more nearly normal S values.  It should, however, be emphasized that the database on which this tentative conclusion was based is still very small. McLaughlin et al. suggest that the S values of the cD galaxies in BM I clusters may have been diluted by mergers with other low S galaxies.  Some support for the view that central cD galaxies may have swallowed numerous low-luminosity galaxies is provided by the recent observations of Barkhouse, Yee & López-Cruz (2000).  Their data show that (a) clusters containing cD galaxies presently have a below-average dwarf-to-giant ratio, and (b) this deficiency is more pronounced in the cluster core than it is far from the central cD galaxy.

Blakeslee (1999) has placed a different spin on the apparent correlation between the S values of central cD galaxies and the BM classes of their parent clusters.  He notes that the S values for central galaxies in rich clusters increase with the velocity dispersion of their parent clusters.  Since velocity dispersion grows with increasing cluster mass, it is therefore not surprising that Blakeslee also find a correlation between S and cluster X-ray luminosity.

A variety of scenarios have been proposed to account for the high S values of many central cluster galaxies. For example (1) West (1993) has proposed that the high specific frequency values that are observed for cD galaxies are due to "biased" globular cluster formation in the rare high-density peaks of the primordial density distribution. On the other hand (2) Hilker, Infante & Richtler (1999) have discussed the possibility that the S values of central galaxies in rich clusters of galaxies might have been enhanced by capture of globular cluster-rich dwarfs such as the Fornax dwarf spheroidal. However, a possible problem with this scenario is that van den Bergh (1998a) finds typical Local Group dSph galaxies (unlike Fornax) to have relatively low S values. An obvious caveat is that early-type dwarfs in rich clusters may have had a different evolutionary history, and cluster formation rate, than those in the Local Group. <u>Hubble Space Telescope</u> observations by Miller et al. (1998) show that the nucleated dE galaxies, which inhabit the core of the Virgo cluster have $S = 6.5 \pm 1.2$, which is not high enough for capture of these objects to raise the S values of cD galaxies to the high values that are observed in some of these objects. Côté, Marzke & West (1998) suggest that the metal-poor cluster population component of giant galaxies like M 87 might have been stripped from less luminous galaxies. (3) The work of Murali (1998) indicates that the high specific cluster frequencies in cD galaxies might, at least in part, be due to the fact that the mean density in these huge galaxies is small, resulting in a low tidal cluster destruction rate, and hence a high present S value. However, the fact that S values are typically about three times greater in cD galaxies than they are in E galaxies would then require that ~ 2/3 of all original globulars have been destroyed in normal elliptical galaxies. Since low-mass star clusters are most easily destroyed by tidal stresses one would expect low-luminosity clusters to be much more depleted than luminous ones. This prediction conflicts with the observation that the globular cluster luminosity function appears to be universal in all environments (Harris 1991). Yet another scenario has been proposed (4) by Forte, Martínez & Muzzio (1982) who suggested that the high S values in luminous central cluster galaxies like M 87 was due to capture of clusters from less massive cluster galaxies. The main problems with this hypothesis are (a) that tidal captures would need to increase the number of globular clusters associated with central cD/E galaxies by a factor of ~ 3, and (b) many of these captured globular clusters would probably be metal-poor. M 87 would have to capture ~ 10,000 globulars to account for its high present S value. It might be difficult to reconcile this stripping requirement with the observation (Harris 1991) that the ellipticals in Virgo (excluding M 87) presently have a higher <S> value than do field ellipticals. Furthermore encounters will result in transfer of both clusters and field stars, i.e. both the luminosity of the parent galaxy and the size of its cluster population might be enhanced by collisions. The fact that the specific cluster frequency is higher in the outer halos of galaxies than it is in the main bodies of such objects [See Fig. 21 of McLaughlin, Harris and Hanes (1994)] will, however, result in some increase in S during mass transfer. This is so because the outer regions of galaxies will be most strongly affected by tides. Muzzio (1987) has emphasized the importance of cluster swapping during tidal encounters. An additional complication (5) is, as Muzzio has emphasized, that tidal interactions will also detach some clusters from galaxies resulting in the formation of a population of intra-cluster tramp globulars that belong to the cluster as a whole. Such clusters are no longer bound to any individual galaxy. This idea was developed in more detail by West et al. (1995), who regard the excess of globular clusters, that appears to be associated with central cD galaxies, as a manifestation of the dense core of an intra-cluster globular cluster population. However, the Coma cluster galaxy IC 4051 (Woodworth & Harris 2000) which has $S = 11 \pm 2$, but a radial velocity that differs from the Coma cluster mean by 1910 km s$^{-1}$, appears to pose a serious problem for this otherwise attractive hypothesis. Radial velocity observations (Kissler-Patig et al. 1999) show that the velocity dispersion of the globular clusters surrounding the E/cD galaxy NGC 1399 increases with radius. This indicates that their dynamics are either determined by the dark matter associated with the parent cluster of NGC 1399, or with the dark matter halo of NGC 1399 itself. [The distinction between these two alternatives may actually be artificial and have no physical significance.] The observation (6) that S(spirals) < S(ellipticals) is, at least in part, due to the fact that the young disk contributes significantly to the total luminosity of spiral galaxies. This suggests that it might be more meaningful to include only the bulge component of the luminosities of spirals in the calculation of their S values. Following this line of argument one might conclude (McLaughlin 1999) that it would be better to look at the number of globular clusters per unit mass. However, a difficulty with this methodology is that dark matter constitutes a large (or dominant) fraction of the mass within the volume occupied by globular clusters. An additional problem with this approach is that M 33 has a significant globular cluster population but little, if any, bulge! It is possible (7) that some special process, such as the existence of cooling flows (Fall & Rees 1985), might have enhanced the rate of globular cluster formation in central cD galaxies clusters. An argument against the latter hypothesis is, however, that the existence of strong cooling flows does not appear to be correlated with high S values. (8) Harris, Harris & McLaughlin (1998) have argued that the high S values of some cD galaxies might be due to strong winds generated by the formation of a luminous cD galaxy. Such winds might have driven away the gas that would otherwise have formed the last generations of galactic field stars. In other words the high S values of some central cD galaxies might be due to a deficiency of field stars, rather than to an excess of globulars. Finally (9) McLaughlin (1999) has argued that the size of the cluster population in a galaxy should, if the cluster formation efficiency is a universal constant, be proportional to the amount of gas that was initially present, i.e. to M(gas) + M(stars). This hypothesis is seemingly contradicted by the observations of the Coma cluster galaxy IC 4051 (Woodworth & Harris 2000), for which X-ray observations indicate that the amount of gas present is insufficient to account for the high specific cluster frequency (S = 11 ± 2) of this galaxy. This problem might be circumvented by making the plausible assumption that the gas that once surrounded IC 4051 was stripped from this galaxy by rampressure as it plunged through the core of the Coma cluster with a velocity of ~ $2 \times 10^3$ km s$^{-1}$. Woodworth & Harris (1999) point out that the discovery of a low S giant elliptical embedded in a massive x-ray halo would contradict their hypothesis.

### 2.3 The high S values for some dE and dSph galaxies

A dwarf galaxy that loses gas after most of its globular clusters have been made, but before the bulk of its field stars have had an opportunity to form, will end up having a high S value. Such gas loss might result from galactic winds, tidal effects (e.g. NGC 4486B) or, more speculatively, as a consequence of the eruption of a nearby active galactic nucleus. NGC 3115B (Durrell et al. 1996b), which is a nucleated dE that has S ≃ 6, compared to S = 2.3 ± 0.5 for its S0(7) parent galaxy NGC 3115 (Hanes & Harris 1986). The fact that NGC 3115 contains a $2 \times 10^9$ M$_\odot$ black hole (Kormendy et al. 1996) suggests that a quasar-like eruption in the nucleus of NGC 3115 might have blown the gas out of both NGC 3115, and its close companion NGC 3115B. In other words the explosion in NGC 3115 could have transformed this galaxy from an early-type spiral to an S0 galaxy. The difference between the S values of these objects might be accounted for by assuming that the explosion in the nucleus of NGC 3115 occurred after the globulars had formed in NGC 3115B, but before it had a chance to form the bulk of its stars. [Since NGC 3115 is a field galaxy, gas depletion by ram-pressure stripping does not appear to be an attractive scenario for its gas depletion]. Kavelaars (1998) and Kundu (1999) have shown that the red globulars in NGC 3115 are associated with the thick disk population component of this galaxy. One might speculate that these red clusters formed during a minor merger that did not destroy the disk of NGC 3115.

Using the Canada-France-Hawaii telescope Durrell et al. (1996a) found little systematic difference between the S values of nucleated and non-nucleated early type dwarf galaxies in the Virgo cluster. However, more recent HST observations by Miller et al. (1998) do appear to show a remarkable difference between the specific cluster frequencies of nucleated and non-nucleated dE galaxies in the Virgo region. According to Miller et al. non-nucleated E galaxies in Virgo typically have low S values. After correction for the effects of luminosity evolution, the specific cluster frequencies in such objects are similar to those in late-type disk galaxies. On the other hand Miller et al. find that the S values of nucleated dE galaxies in the Virgo cluster depend on luminosity, with the highest S values occurring in the faintest nucleate dE systems. Possibly the cluster environment (or rampressure stripping) curtailed the formation of second generation stars in the main bodies of these objects after the first generation of stars and globular clusters had formed. As has previously been noted by van den Bergh (1986), and by Ferguson & Sandage (1989), nucleated Virgo dE galaxies have a distributed that is strongly concentrated towards the core of the Virgo cluster. In this respect their distribution resembles that of the giant E galaxies in the Virgo cluster. On the other hand non-nucleated dE's have a more diffuse spatial distribution, which resembles that of late-type disk galaxies.

The Fornax dwarf spheroidal galaxy, which is a distant companion to the Milky Way system, also has an unusually high specific cluster frequency S ~ 29. The reason for this is presently not known. From the observation that none of the faintest dwarf spheroidal companions to the Galaxy have any globulars it may be concluded (van den Bergh 1998a) that the dwarf spheroidals fainter than $M_V =$ -12 typically do not share the high S value observed for Fornax. The very high specific globular cluster frequency in the Fornax dwarf spheroidal presents a formidable challenge to theories (e.g. Kumai, Basu & Fugimoto 1993, Manning 1999) of globular cluster formation. In the scenario proposed by Manning globular clusters are formed when Ly $\alpha$ clouds crash into the gas surrounding a massive attractor. He points out that this process will not work for objects that are unable to sustain circular velocities of at least ~ 40 km s$^{-1}$, i.e. for low-mass objects such as the Fornax dwarf in which the central velocity dispersion is only 10.5 km s$^{-1}$ (Mateo 1998). In addition to Fornax the Sagittarius dwarf also appears to have an above-average S value, although its exact value remains uncertain because we do not know how much mass (luminosity) has been tidally stripped from the Sagittarius system. In Virgo, Durrell et al. (1996a) also find high S values for the nucleated dwarf ellipticals VCC 1311 and VCC 1333 which have S = 15 and S = 12, respectively. McLaughlin (1999) suggests that the high S values of the lowest luminosity dSph galaxies are due to the fact that these objects were not able to retain gas in their shallow potential wells. However, it appears unlikely that gas loss can account for the very high S value of the Fornax dwarf. This is so because the faintest Galactic dwarf spheroidals, which would be expected to have lost even more gas than Fornax, collectively have S < 5 (van den Bergh 1998a) - which is significantly lower than the value S ~ 29 for the Fornax dwarf.

Assuming S = 5 one would have expected the Local Group dwarf elliptical M 32 to contain ~ 20 globular clusters. In fact, none are observed. This suggests that some of the relatively metal-poor globulars, that are presently situated in the halo of M 31, may represent objects that were tidally stripped from the outer regions of M 32. Furthermore some globular clusters, that were initially in the core of M 32, might have been dragged into its nucleus by tidal friction (Tremaine, Ostriker & Spitzer 1975). For the same reasons one might expect the tidally truncated galaxy NGC 4486B (Faber 1973, Forbes, Brodie & Grillmair 1997b) to have a below-average value of S. This expectation could be checked by imaging this object with the <u>Hubble</u> <u>Space</u> <u>Telescope</u>.

### 3. COLOR DISTRIBUTION OF GLOBULAR CLUSTERS

It was first shown by van den Bergh (1975) that the mean colors of globular clusters are redder in systems with luminous parent galaxies than they are for the globular cluster systems of less luminous galaxies. Color distributions within individual globular clusters in systems surrounding 50 (mostly early-type) galaxies have recently been published by Gebhardt & Kissler-Patig (1999). Roughly half of these systems show evidence for bimodality in the distribution of their globular cluster colors. (Bimodality in some globular cluster systems may have been hidden by photometric errors). Most luminous elliptical galaxies (Neilsen & Tsvetanov 1999) are observed to have bimodal color (metallicity) distributions. However, an exception is the luminous S0/E3 Virgo galaxy M86 (NGC 4406), which has only a single (blue) peak in the color distribution of its globular clusters. A counter-example is provided by the Coma elliptical IC 4051 (Woodworth & Harris 2000) which appears to lack a significant blue (metal-poor) cluster population. This would suggest that either (a) this E galaxy did not start forming globulars until the metallicity in its gas had reached quite a high level, or (b) that the outer blue metal-poor cluster population was tidally stripped from IC 4051. However, the fact that no blue clusters are seen at small galactocentric distances militates against this hypothesis. Kundu (1999) also finds that the E1 galaxy NGC 7626, and the S0 galaxy NGC 524, both appear to have only a single (red) cluster population. Another example of an object with an unimodal color (metallicity) distribution (which peaks at [Fe/H] $\simeq$ -1) is the isolated S0 galaxy NGC 7457 (Chapelon et al. 1999). This relatively high metallicity suggests that NGC 7457 does not contain a significant blue cluster population. Perhaps this rather isolated object did not merge with [or strip clusters from (Côté, Marzke & West 1998)] any low-luminosity fragments that contained metal-poor blue globulars. Harris, Harris & Poole (1999) suggest that the metal-poor globular cluster population component surrounding the peculiar E galaxy NGC 5128 formed while its protogalaxy was still in a clumpy and fragmented state, leaving most of the gas unused. On the other hand the second and larger burst of cluster formation in this galaxy appears to have taken place when (by now enriched) gas had collected into the fully formed potential well of the elliptical galaxy NGC 5128. In summary it appears that a wide variety of cluster forming scenarios needs to be invoked to account for both the range in specific cluster frequencies S, and for the observed differences in the color distribution of globular cluster systems.

Forbes, Brodie & Grillmair (1997a) have shown that the mean colors of the red (metal-rich) population component of globular clusters become redder as the luminosities of their parent galaxies increase. This shows that the evolution of these red globulars is intimately tied to that of their parent galaxy. On the other hand, the average colors of the blue component of the globular cluster population of globular clusters appear to be independent of the luminosity of their parent galaxy. Taken at face value this result suggests that these metal-poor blue globulars were drawn from a primordial population that filled the galaxy cluster in which each of these parental galaxies was formed (West et al. 1995). Such an argument cannot be excluded yet for elliptical galaxies in rich clusters. However, three lines of evidence suggest that the blue (metal-poor) population component of the globular cluster systems associated with Local Group galaxies was not derived from such a common pool: (1) The metal-poor globular clusters surrounding the LMC are systematically more flattened than those located in the halo of the Milky Way system. (2) The globular clusters in both the LMC and in the Galactic halo exhibit a strong correlation between their half-light radii $R_h$ and their galactocentric distances $R_{Gc}$ (van den Bergh 1994). (3) Color-magnitude diagrams show that the metal-poor globular clusters in the Galactic halo have systematically different horizontal branch population gradients from those in LMC globular clusters. These observations clearly show that the metal-poor Local Group globular clusters were formed more-or-less in situ, and were therefore not derived from a primordial population of Local Group globular clusters. In this respect, the metal-poor clusters surrounding Local Group spirals may differ from those of the blue globulars that appear to be associated with massive ellipticals in rich clusters.

## 4. MASS SPECTRA OF GLOBULAR CLUSTERS

### 4.1 The shape of the cluster luminosity function

Harris & Pudritz (1994) have speculated that the mass spectrum of massive clusters is related to the mass spectra of the dense gas cores that are scattered within molecular clouds from which they are presumed to have formed. In this connection one question that presents itself insistently is (Fritze-von Alvensleben 1999): Would one expect the same mass spectrum of cloud cores for the molecular clouds of normal quiescent spirals, and in violently interacting objects like The Antennae? By the same token one might ask if it is reasonable to expect the present mass spectra of Galactic cloud cores to be similar to those of the molecular clouds that existed in violently colliding and/or collapsing protogalaxies ~ 13 Gyr ago.

### 4.2 Dependence on environment

The fact that the mass spectra of the red (metal-rich) and blue (metal-poor) globular clusters are almost indistinguishable (Kundu 1999) shows that the process by which globulars formed must have been quite insensitive to metallicity. Compared to halo field stars, the frequency distribution of metallicity among Galactic globulars appears to be deficient in objects with [Fe/H] < -2.5. If the reality of this difference should be confirmed by future work, then it might hint at a below-average globular cluster formation efficiency at very low metallicities. It is curious, and presently not understood, why neither the integrated mass spectra of globular clusters (Harris 1991), nor the mass spectra with which individual stars are formed within clusters and dwarf spheroidal galaxies (Wyse et al. 1999), appears to depend on environment.

The luminosity functions, and hence the mass spectra of star formation, are (at least above $M_V = +8$) indistinguishable in the Ursa Major dwarf spheroidal galaxy and in the globular cluster M 15 (Wyse et al. 1999). This shows that the mass spectrum of star formation is unaffected by density differences of three orders of magnitude, and of the presence (UMi), or absence (M 15), of dark matter.

### 4.3 Radial variations in the globular cluster luminosity function

It is a generic prediction of models for the evolution of cluster systems (Gnedin & Ostriker 1997, Murali & Weinberg 1997) that low-mass clusters, and any but the most massive clusters at small galactocentric distances, will be more easily destroyed than high- mass clusters at large galactocentric distances. As a result one would expect the luminosity distribution of clusters at small galactocentric distances to be more depleted in low-mass clusters than the luminosity distribution of clusters at large galactocentric distances. However, this prediction is strongly contradicted by presently available observations. From ground-based data, that extend out to a radial distance of 30 kpc, Harris, Harris &

McLaughlin (1998) find no evidence for a radial variation in the luminosity distribution of the bright end of the globular cluster luminosity function of the M 87 globulars. Furthermore HST observations of 1057 globulars in the inner region of M 87 by Kundu et al. (1999) provide no evidence for significant radial variations in the cluster luminosity distribution in bins with mean radii ranging from 1.6 kpc to 6.4 kpc (distance of 16 Mpc assumed). A likely explanation for this apparent discrepancy is that the mean radii of globular clusters increase with galactocentric distance. Kundu et al. (1999) find that the half-light radii of blue (outer halo?) globulars are $\approx$ 20% greater than those of the red (inner halo?) globulars. The change in cluster radii with galactocentric distance (van den Bergh 1994) will counteract the expected preferential depletion on the least massive clusters at small galactocentric radii. Furthermore Gnedin, Lee & Ostriker (1999) find that the effects of tidal shocks are self-limiting because the influence of cluster shocking will diminish as the cluster loses mass and becomes more compact.

It is generally assumed that the present log-normal magnitude distribution of globular clusters evolved from an initial power-law luminosity distribution by preferential destruction of the least massive clusters. As pointed out above, a possible problem with this hypothesis is that the predicted radial variation of the cluster luminosity function is not observed. Alternatively it might be assumed that the great burst of globular cluster formation that occurred ~ 13 Gyr ago produced objects with a log-normal mass spectrum. Vesparini (1998) has shown that such an initially log-normal mass spectrum would be conserved over a Hubble time, even when ~ 50% of the original globular cluster population is destroyed. Preliminary evidence in favor of the view that clusters formed in colliding spirals have a log-normal mass spectrum has recently been published by Fritze-von Alvensleben (1999).

**5.    GALAXY MERGERS AND CLUSTER FORMATION**

Toomre (1977) wrote "It seems almost inconceivable that there wasn't a great deal of merging of sizable bits and pieces (including quite a few lesser galaxies) early in the career of every major galaxy...I really do suspect that the few mergers we are still privileged to witness are just the statistical dregs of a once very common process." This view, which is strongly supported by observations in the Hubble Deep Field (van den Bergh et al. 1996), has led to a real paradigm shift in the way we view galaxy evolution. In particular Schweizer (1987) argued that globular clusters might form during collisions, and subsequent mergers, between gas-rich spiral galaxies. Strong confirmation of the idea that galaxy mergers can produce a burst of cluster formation was provided by Fritze-von Alvensleben (1999) and by Whitmore et al. (1999), and references therein, who observed the numerous luminous star clusters and associations in the violently interacting spirals NGC 4038/4039 (The Antennae) with the Hubble

Space Telescope. It would be important to obtain spectra of the "young globular clusters" in NGC 4038/39 to see if they exhibit the same peculiar stellar luminosity distribution that appears to occur in the "young globular clusters" in NGC 1275 (Brodie et al. 1998). If the young populous clusters in NGC 4038/39 also lack low-mass stars they would not survive for a Hubble time to form true globular cluster-like objects.

A detailed discussion of evidence in favor of the merger picture from globular cluster data is given in Zepf & Ashman (1993) and by Ashman & Zepf (1998, pp. 115-123), while some of the difficulties with this scenario are highlighted by Forbes, Brodie & Grillmair (1997) and Forbes (1999). These authors point out that high S galaxies have an above-average fraction of metal-poor clusters, whereas one would have expected most of the extra globulars formed by mergers at late times to be metal-rich. This suggests that the excess clusters in high-S galaxies might, in fact, be members of the surrounding galaxy cluster core. This idea is also supported by the fact that these metal-poor clusters are preferentially located in the outer regions of their parent galaxies. Forbes et al. speculate that the metal-poor globulars were typically formed at an early stage in the collapse of their ancestral galaxy, whereas the metal-rich ones were created at the same time as the bulk of the stars in ellipticals. If this interpretation is correct, than the metal-rich globulars in ellipticals are the analogs to the metal-poor clusters in spirals! On this picture the disk globulars would be objects that formed during a third phase of star and cluster formation.

An unexpected challenge to the hypothesis that the "young globular clusters" in galaxies were produced by mergers is provided by the puzzling observations of Brodie et al. (1998), who find that the spectra of these objects in NGC 1275 do not resemble those of young populous clusters in the Magellanic Clouds. Their Balmer lines are seen to be too strong to be consistent with the evolution of clusters having a standard mass distribution. The spectral energy distributions of five such objects in NGC 1275, which they obtained with the Keck telescope, appear to require that these clusters have a truncated initial stellar mass function with a mass range of 2-3 $M_\odot$ and an age of 450 Myr. If this interpretation is correct then these objects would fade rapidly after the 2 $M_\upsilon$ stars have evolved and died. Taken at face value this result would imply that the "young globulars" found in the core of NGC 1275 cannot be the ancestors of typical globular clusters. It would clearly be very important to obtain similar spectral observations of "young globular clusters" in merging spirals, such as NGC 4038/39 to see if they might also suffer from the same small mass-range syndrome.

**6.     DIFFERENCES BETWEEN GLOBULAR CLUSTERS AND DWARF SPHEROIDALS**

Globular clusters and dwarf spheroidal galaxies have similar metallicities, ages, integrated luminosities and stellar mass spectra.  However, they do exhibit three significant differences that must be deeply rooted to their evolutionary histories:  (1) Most (or all) dwarf spheroidals contain missing mass (Mateo 1998), whereas no evidence for any dark matter has ever been found in a globular cluster.  (2) The stars in dwarf spheroidals exhibit a significant range in metallicity, but globular cluster stars typically have very similar [Fe/H] values[1].

---
[1]     Omega Centauri, which is the most luminous Galactic globular cluster exhibits a range of metallicities (Lee et al. 1999), suggesting that it may have been able to form second generation stars by hanging on to some of the metal-enriched gas ejected by evolving first generation stars.

---

The only well-established exception is ω Centauri which contains both an old metal-poor population and a more metal-rich centrally concentrated population that is ~ 3 Gyr younger (Hughes & Wallerstein 1999, Lee et al. 1999, Hilker & Richtler 1999).   Majewski et al. (1999) have argued that both the orbit of, and the metallicity dispersion in, ω Cen suggest that this object is the nucleus of a now defunct dwarf spheroidal galaxy.  (3) The stellar densities in globular clusters are more than three orders of magnitude greater than those in typical globulars.

It is of interest to note that both the diameters of globular clusters (Kundu 1999) and their luminosities (Harris 1991) appear to be very similar, even when their parent galaxies are of quite different luminosity or Hubble type. However, the mean ellipticity of globulars does appear to differ significantly from galaxy to galaxy. For example the globulars in both the Magellanic Clouds (van den Bergh 1982b) and in NGC 5128 (Holland, Côté & Hesser 1999) appear, on average, to be much more flattened than they are in M 31 the Milky Way system. In this connection it is noted that the young star clusters in the LMC are also more flattened than those in the Galaxy. The reasons for such galaxy-to-galaxy differences in cluster flattening remain entirely unknown.

## 7. SYSTEMATICS OF GLOBULAR CLUSTER SYSTEMS

Extensive information on richness and structure of over 50 globular cluster systems is presently available (van den Bergh 1998b). Since late-type galaxies tend to have low S values their cluster systems are generally not very populous. As a result it is not yet possible to draw statistically significant conclusions about the global characteristics of the cluster systems surrounding galaxies of types Sb, Sc and Ir. However, the situation is more favorable for the richer cluster systems surrounding early-type galaxies. It is found that most luminous early-type galaxies are embedded in cluster systems with shallow radial density gradients. Furthermore the color (metallicity) distributions of the globular cluster systems surrounding luminous parent galaxies are mostly seen to be broader than those which are associated with less luminous galaxies. Taken at face value this suggests that the cluster systems surrounding less luminous parent galaxies may have had simpler evolutionary histories than those associated with luminous parent galaxies. However, a problem with the interpretation of these data is that it is difficult to disentangle the strongly correlated effects of high parent galaxy luminosity, presence of a core or boxy isophotes, and radial cluster density gradients.

## 8. CONCLUSIONS

Perhaps the strongest conclusion to be drawn from the rapidly growing volume of data on the globular cluster systems of galaxies is that they have undergone a bewildering variety of evolutionary histories. Even after evolutionary dimming is taken into account, the galaxy-to-galaxy variations in the specific globular cluster frequency are seen to span an order of magnitude. Even larger differences are observed between the efficiencies with which massive young clusters are formed in different galaxies. Within individual systems the fraction of all star creation that results in globular cluster formation is found to vary with radius and time. Presently available data suggest, but do not yet prove, that cluster formation in the Large Magellanic Cloud (van den Bergh 1999a) was favored in regions of violent chaotic star formation. This may indicate that strong shocks can trigger formation of massive open clusters. The great burst of globular cluster formation in the Galaxy and the LMC, which took place ~ 13 Gyr ago, may have been triggered by the strong shocks that accompanied the initial collapse of these galaxies. It is not yet clear why this burst of cluster formation stopped so precipitously in the Galaxy and the LMC, but continued for a few Gyr in dwarf spheroidal galaxies. Nor is it understood why cluster formation turned on rather gradually in the Small Magellanic Cloud.

It is particularly puzzling that globular cluster formation in M 33 appears to have taken place in the halo of the Triangulum galaxy, while the globulars in the Large Cloud did not form until that system had collapsed into a disk. Within the present paradigm of galaxy evolution it is also very difficult to understand why the M 87 globular cluster system is significantly flattened, whereas the main body of this E0 galaxy exhibits a spherical distribution of stars. The high specific globular cluster frequency in the Fornax dwarf spheroidal, and in the non-central E luminous galaxy IC 4051, present particularly strong challenges to most popular scenarios for the globular cluster formation. Observations of the highly flattened S0 galaxy NGC 3115, and its companion NGC 3115B, suggest that the formation of stars and globular clusters may suddenly be truncated by the emergence of a quasar. The absence of globular clusters in M 32, even though ~ 20 are expected, shows that the specific globular cluster frequency may depend on peculiarities of individual evolutionary histories. It would be important to obtain spectra of the luminous young star clusters in NGC 4038/39 (The Antennae) to see if these clusters have normal stellar mass spectra, which would allow the most massive clusters to survive for a Hubble time.

It is presently not understood why globular clusters are so similar in parent galaxies that have a wide range in Hubble type, mass, density, and metallicity. Nor is it yet clear why globular cluster luminosity functions appear to be (almost) independent of galactocentric distance. Possibly this effect is due to the fact that globular clusters at small distances, which are subject to the greatest tidal stresses, are also more compact and tightly bound than those at larger galactocentric distances. Another mystery is why globular clusters in the Magellanic Clouds are, on average, much more flattened than their counterparts in M 31 and the Milky Way system. This difference, in conjunction with differences in cluster age distributions, and the fact that the LMC and the Galaxy exhibit a correlation between $R_{Gc}$ and $R_h$, shows that the metal-poor globular clusters in the Local Group were formed more-or-less in situ. In other words they did not derive from a single primordial reservoir of metal-poor Local Group globular clusters. In this respect, the metal-poor globulars associated with Local Group spirals might differ from the metal-poor clusters surrounding elliptical galaxies in rich clusters. Finally it is noted that both metal-rich and metal-poor globular clusters have group characteristics that correlate with the luminosities of their parent galaxies. This shows that the "family traits" of globular clusters depend, to some extent, on the properties of their parent galaxies.

It is a pleasure to thank Roberto Capuzzo-Dolcetta, Pat Côté, Bruce Elmegreen, Duncan Forbes, Satoru Ikeuchi, Richard Larson, Arunav Kundu, Dean McLaughlin, Chigurupati Murali, George Preston, Ralph Pudritz, Mario Vietri and Michael West for useful exchanges of e-mail. I also thank referee Bill Harris for numerous helpful suggestions.